\documentclass{aa}
\usepackage{amssymb}
\usepackage{graphicx}
\begin{document}
\thesaurus{08(08.02.6,08.16.5,08.09.1,08.12.1,08.06.2)}
\offprints{G. Duch\^ene}
\title{Low-mass binaries in the young cluster IC\,348: implications for
  binary formation and evolution\thanks{Based on
    observations made with the Canada-France-Hawaii Telescope, operated by the
    National Research Council of Canada, the Centre National de la Recherche
    Scientifique de France and the University of Hawaii}}
\author{G. Duch\^ene\inst{1}, J. Bouvier\inst{1} \and T. Simon\inst{2}}
\institute{Laboratoire d'Astrophysique, Observatoire de Grenoble, Universit\'e
    Joseph Fourier, BP 53, 38041 Grenoble Cedex 9, France,
    Gaspard.Duchene@obs.ujf-grenoble.fr,
    Jerome.Bouvier@obs.ujf-grenoble.fr
\and Institute for Astronomy, University of Hawaii, 2680, Woodlawn Drive,
    Honolulu, HI 96822, U.S.A., simon@ifa.hawaii.edu}
\date{Received 17 November 1998; accepted 4 January 1999}
\titlerunning{Low-mass binaries in the young cluster IC\,348}
\maketitle

\begin{abstract}
  
  We report on a near-infrared adaptive optics survey of a sample of
  66 low-mass members of the pre-main sequence stellar cluster
  IC\,348. We find 12 binary systems in the separation range
  $0\farcs1$--8$''$, excluding 3 probable background projected
  companions. An estimate of the number of faint undetected companions
  is derived, before we evaluate the binary frequency in this cluster.
  In the range $\log P=$\,5.0--7.9\,days, the binary fraction in IC\,348
  is $19\pm5$ \%. This is similar to the values corresponding to G-
  and M-dwarfs in the solar neighbourhood population ($23\pm3$ \% and
  $\sim18$\%, respectively). Furthermore, the distribution of orbital
  periods of IC\,348 binaries in this range is consistent with that of
  field binaries. We conclude that there is no binary excess in
  IC\,348.
  
  Substellar companions are found to be rare, or even missing, as
  companions of low-mass stars in the separation range we surveyed.
  Also, the mass ratio distribution is not peaked at $q\approx1$ in
  IC\,348, and it is unlikely that an observational bias can account
  for that. 

  We do not find any evidence for an evolution of the binary frequency
  with age within the age spread of the cluster of about 10\,Myr.
  Comparing the binary frequency in IC\,348 with that of other star
  forming regions (SFRs) and young open clusters, we conclude that
  there is no significant temporal evolution of the binary fraction
  between a few Myrs after the formation process and the zero-age main
  sequence (ZAMS) and field populations. We find instead a trend for
  the binary fraction to be inversely correlated with stellar density,
  with dense clusters having a binary fraction similar to that of
  field dwarfs and loose associations exhibiting an excess of
  binaries. Two scenarios can be suggested to explain these
  differences: either all SFRs, clusters and associations alike,
  initially host a large number of binaries, which is subsequently
  reduced only in dense clusters {\it on a timescale of less than 1
    Myr} due to numerous gravitational encounters, or specific initial
  conditions in the parental molecular clouds impact on the
  fragmentation process leading to intrinsically different binary
  fractions from one SFR to the other.

\keywords{binaries: visual -- stars: pre-main sequence -- stars: imaging --
  stars: late-type -- stars: formation}

\end{abstract}

\section{Introduction}

Several studies in the early 90s have shown that binarity is a very
common property of low-mass main sequence stars: about 53\% of G-type
stars, 45\% of K-dwarfs and 42\% of M-dwarfs are in fact multiple in
the solar neighbourhood (Duquennoy \& Mayor \cite{duq_may:1991},
hereafter DM91; Mayor et al. \cite{mayor:1992}; Fischer \& Marcy
\cite{fischer_marcy:1992}). An important issue for current star
formation models is to account for the high number of binaries, and to
predict their physical properties.

The relative number of binary systems may be even larger among much
younger stellar populations. One of the best studied low-mass SFRs,
the Taurus-Auriga dark cloud, hosts almost twice as many binaries as
the solar neighbourhood in the separation range 2--2000\,AU (Leinert
et al. \cite{leinert:1993}; Ghez et al. \cite{ghez:1993}; Simon et al.
\cite{simon:1995}). Yet, subsequent surveys of a number of other SFRs
have led to somewhat conflicting results: while some exhibit binary
excesses comparable to that of the Taurus cloud (Padgett et al.
\cite{padgett:1997}; Ghez et al. \cite{ghez:1997}), others appear to
have similar binary fractions as field dwarfs (Brandner et al.
\cite{brandner:1996}. For instance, the Orion Trapezium shows a binary
fraction in good agreement with that on the MS (Petr et al.
\cite{petr:1998}; Prosser et al. \cite{prosser:1994}). Duch\^ene
(\cite{duchene:1998bin}) recently reanalysed in a consistent way these
various studies and confirmed that the binary fraction appears to vary
from one SFR to the other, with the main exception of all Orion
clusters whose binary fraction is similar to that of the field.

Several proposals have been made to account for these results. It has
been suggested that the fragmentation process during the protostellar
collapse yields a high fraction of multiple systems which, however,
steadily declines over time as multiple systems are disrupted during
their subsequent evolution (Ghez et al. \cite{ghez:1993}). Then, the
binary fraction would depend on the age of the stars and would vary
over a timescale of several 100\,Myr (Patience et al.
\cite{patience:1998}). Alternatively, it is conceivable that the
binary fraction of a cloud is established at the very beginning of the
cluster history.  Kroupa (\cite{kroupa:1995a,kroupa:1995b}) has
recently shown from $N$-body simulations that in regions as dense as
the Trapezium cluster the binary fraction could decrease from 100\% to
about 50\% in less than 1\,Myr due to gravitational encounters. Still
another possibility is that the binary frequency is sensitive to
environmental conditions in the parental molecular cloud. In a
qualitative study, Durisen \& Sterzick (\cite{durisen_sterzik:1994}) found
that both the current fragmentation models and disk instabilities are
compatible with a lower binary fraction in clouds with higher
temperatures.

In order to distinguish between these alternatives, we started a
long-term project aimed at studying binaries in clusters at different
evolutionary stages, from the birthline to the MS. In Bouvier et al.
(\cite{bouvier:1997}), we already found that the binary frequency in
the 100\,Myr old Pleiades cluster is similar to that of the MS. We
report here the results obtained in the pre-main sequence (PMS)
cluster IC\,348. This cluster was selected on the basis of its age
being similar to that of the Taurus cloud (about 2\,Myr), but its
stellar density being much larger (about 500\,stars\,pc$^{-3}$
compared to a few stars\,pc$^{-3}$ for Taurus).

IC\,348 is a young cluster located in the Perseus molecular cloud, at
a distance of about 320 pc (Herbig \cite{herbig:1998}, hereafter H98).
It hosts a B5 V star (BD+31$^\circ$643), about 100 optical sources
(Trullols \& Jordi \cite{trullols_jordi:1997}), as well as a few
hundred infrared sources, which are probably embedded young stars
(Lada \& Lada \cite{lada_lada:1995}). The age of this cluster has been
estimated by several methods: Lada \& Lada, by fitting the IMF,
estimate that star formation is still going on after a burst
$\sim$5--7\,Myr ago; Luhman et al. (\cite{luhman:1998}), hereafter
L98, found that a major burst occured $\sim3$\,Myr ago, but that stars
as old as 10\,Myr also lie in the cluster. Similarly, from a
dereddened colour-magnitude diagram, H98 estimates ages ranging from
less than 1\,Myr and up to about 10\,Myr for about 100 members, with a
median age of $\sim2$\,Myr. From a deep near-infrared survey, Lada \&
Lada estimated a stellar density of about 500 stars\,pc$^{-3}$ or
220$M_\odot$\,pc$^{-3}$ (within the half-mass radius of 0.47\,pc), and
a projected surface density of about 1000 stars\,pc$^{-2}$ in the
central 0.1\,pc, similar to the NGC 2024 cluster in Orion. H98 also
evaluated masses for the members, and found that the median mass is
about 0.5$M_\odot$, in agreement with the IMF estimated for this
cluster (which is very similar to the IMF from Scalo
\cite{scalo:1986}). Furthermore, H98 conducted an H$\alpha$ survey,
and discovered over 110 emission-line stars in a $7\farcm5\times15'$
area centred on the cluster; all of these stars are very likely to be
young, active cluster members; about 70 of them have independently
been confirmed as members on the basis of colour-magnitude diagrams.
Preibisch et al. (\cite{preibisch:1996}) performed a systematic X-ray
survey of the area, detecting over 110 sources within a
1$^\circ$-radius circle.

We report on our adaptive optics observations of IC\,348 in
Sect.\,\ref{obs}, and estimate the binary fraction in the cluster in
Sect.\,\ref{fraction}. In Sect.\,\ref{properties} we discuss various
other binary properties (especially the scarcity of brown dwarf
companions in our survey), and Sect.\,\ref{discus} presents a discussion
on the link between the binary fraction and environmental conditions
in SFRs. Sect.\,\ref{concl} summarizes our conclusions.

\section{Observations}
\label{obs}

Our sample was selected from H98's list: all stars brighter than
$R\approx15$ were observed, with a 75\% completeness level at
$R=16.5$; overall, the survey is two-third complete at the $H=12$
limit. Mosaicing with small offsets from the brightest stars, about 25
additional fainter stars were surveyed. In total, we observed 70 of
H98's members (within 66 independent systems); we also observed 24
stars for which H98 could not assess membership. Ten of these stars
are considered as members by L98, on the basis of the detection of the
Li $\lambda6707$\,\AA\, absorption line and of spectral classification
(for late M stars).

\begin{figure*}[t]
\begin{center}
\includegraphics[width=\textwidth]{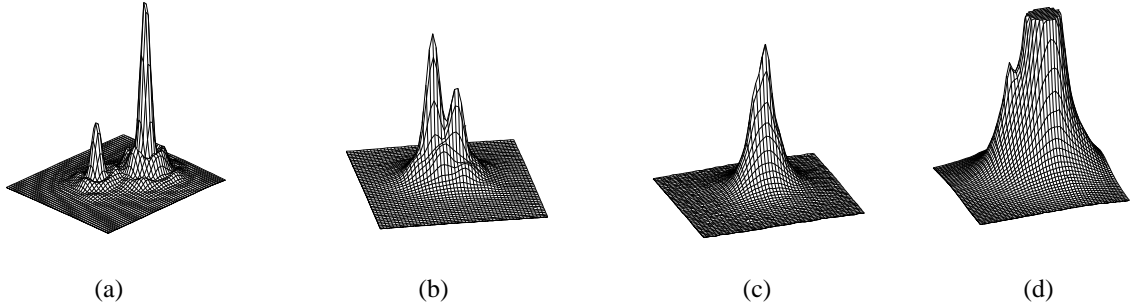}
\end{center}
\caption{Examples of images obtained in our survey. Each box is
  $1\farcs75\times1\farcs75$. ({\bf a}):
  IfA\,166 is a bright star ($R=11.2$), allowing diffraction-limited
  images, with high Strehl ratios at $K$ -- ({\bf b}): the fainter
  ($R=15.8$) binary IfA\,119 is still resolved at $K$ with a
  separation of $0\farcs25$ -- ({\bf c}) IfA\,192 is too faint
  ($R=16.5$) to be resolved as a binary (separation of $0\farcs13$),
  but it is clearly elongated at $J$, $H$ and $K$, and a deconvolution
  process was used to obtain the relative photometry -- ({\bf d}):
  IfA\,184 illustrates the limit for the detection of close, faint
  companions (the image is in the $J$ band).}\label{example}
\end{figure*}

The observations were obtained during four nights in December 1997 at
the Canada-France-Hawaii Telescope on Mauna Kea. We performed
near-infrared ($JHK$), high angular resolution imaging using the
Adaptive Optics Bonnette and the new infrared camera KIR, a
$1024\times1024$ HAWAII detector. The pixel scale is
$0\farcs0351$/pix, yielding a total field-of-view of $36''\times36''$.
Most of the images are diffraction-limited in $H$ and $K$, but the
images of the faintest stars or those observed at large offsets from
the wavefront star ($\gtrsim20''$), have FWHM as large as $0\farcs2$;
Fig.\,\ref{example} illustrates the image quality in our survey. We
surveyed each target in $H$, and all resolved systems were also
observed in $J$ and $K$ (with the exception of pairs formed by two
stars already known as members). A typical observing sequence consists
of 16 images, at 4 positions, with individual exposure times ranging
from 2 to 30 seconds to avoid saturation of the sources. On the first
night, the observing procedure was different, since we produced
$1\farcm5\times1\farcm5$ mosaics centred on the brightest stars close
to the cluster centre. In these images, the exposure time was such
that no star was saturated; the 5\,$\sigma$ detection limit in these
images is about $H=15.5$. For all but two H98 members (IfA\,134 and
163), this leads to a detection limit of at least $\approx3$\,mag for
separations larger than $0\farcs5$.

\begin{figure}[t]
\begin{center}
\includegraphics[height=\columnwidth,angle=270]{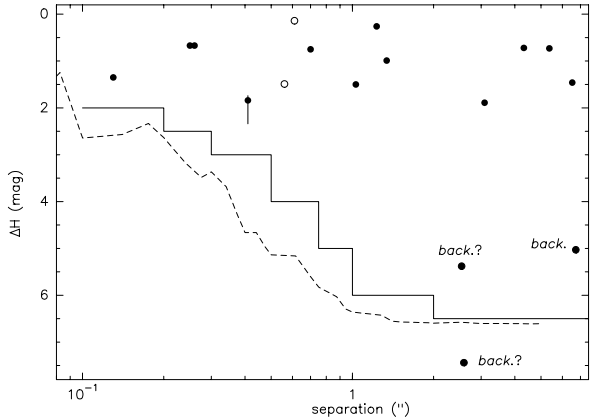}
\end{center}
\caption{The observed binaries in our sample, including the probable
  background stars (labelled "back."). Open circles represent L98's
  members detected as binaries. The error bars are smaller than the
  symbol size, except for IfA\, 184 (see text). The dashed line is an
  estimate of the detection limit: it is the 3\,$\sigma$ noise level
  in the averaged radial profile of a single star. The histogram is
  another estimate of this limit, obtained by artificially adding
  faint companions close to single stars.}\label{limit}
\end{figure}

UKIRT photometric standards were observed every night for the flux
calibration. Data reduction was performed with IRAF packages.
Astrometry and relative photometry of binaries were obtained by PSF
fitting and then combined with large-aperture photometry to get the
absolute photometry of each component. In two cases (IfA\,100, 102)
deconvolution had to be applied to obtain the relative photometry; we
used the Iterative Deconvolution Analysis in C (IDAC) routine
(Jefferies \& Christou \cite{jefferies_christou:1993}), and
cross-checked with the Lucy algorithm. In the case of IfA\,184,
deconvolution was unsuccessful in the $H$ band, and we consider here
the results from PSF fitting, although we think that this method leads
to an underestimate of the flux ratio in this specific case. Estimates
of the uncertainties are 0.05\,mag and 0.02\,mag for absolute and
relative photometry, $0\farcs005$ for the separation and $0\fdg2$ for
the position angle. The errors are slightly larger when deconvolution
was applied; when the adaptive optics system was locked on a binary
system (e.g., IfA\,139--140), the PSF was substantially deformed,
leading to increased uncertainties on the centroid locations.

\section{Binary fraction in IC\,348}
\label{fraction}

Our binary candidates are listed in Table\,\ref{photom}, together with
their astrometric and photometric measurements. To accept two stars as
a binary candidate, we have set an upper limit for the separation of
8$''$, corresponding to $\sim$2500 AU. This limit should avoid
confusion between real companions and background stars (see discussion
below). A case-by-case study has been done, however, for each binary
candidate. All the stars that appeared as singles are listed in
Table\,\ref{singles}. The astrometry for the binaries detected by H98
in the $I$ band agrees with ours to within $0\farcs15$ and 2$^\circ$,
even for the closest pairs (IfA\,136 and 211). We failed to detect the
faint companion $1\farcs3$ away from IfA\,159, because the latter is a
faint star ($R=18.7$, $H=12.19$), which was observed in a mosaic on
the first night, with a more limited dynamic range compared to that
usually achieved on other stars. At the 5\,$\sigma$ level, the
companion is fainter than $J=16.2$ and $H=15.2$\,mag or, equivalently,
the binary flux ratio is larger than 2.9\,mag in both bands.

\begin{table*}[t]
\caption{\label{photom} Astrometry and photometry for binary
  candidates in IC\,348. Position angles are given East from
  North. Boldface entries in the first column are H98's members. The
  primary mass is from Herbig (priv. comm.), and takes into account
  individual ages; the mass-ratio is estimated from $\Delta H$ and the
  mass-luminosity relation $M_H=-3.25\log (M/M_\odot) +2.19$ for
  $M<1M_\odot$, Baraffe et al. (\cite{baraffe:1998}). Both methods are
  somewhat uncertain but are independent. {\it TJ\,81\/}=BD+31$^o$643
  does not have an IfA number, but is designed by its
  Trullos \& Jordi (\cite{trullols_jordi:1997}) identification.}
\begin{tabular}{l|lll|lll|ll|cc}
\hline
 IfA & $J_{tot}$ & $H_{tot}$ & $K_{tot}$ & $\Delta J$ & $\Delta
 H$ & $\Delta K$ & $\rho('')$ & P.A. ($^\circ$) & $M_A (M_\odot)$ &
 $M_B/M_A$ \\
\hline
\multicolumn{11}{c}{Probable member binaries}\\
\hline
 {\bf 48}--{\bf 49} & & 12.41 & & & 0.99 & & 1.343 & 112.4 & 0.2 &
 0.50 \\
 {\bf 85}--82 & & 11.70 & & & 1.89 & & 3.086 & 289.2 & 0.4 & 0.26 \\
 {\bf 102}$^1$ & 12.29 & 11.25 & 11.10 & 0.69$\pm$0.05 & 0.67$\pm$0.05 &
 0.62$\pm$0.05 & 0.26$\pm$0.01 & 357.0$\pm$0.5 & 0.4 & 0.62 \\
 {\bf 119} & 12.16 & 11.30 & 11.00 & 0.63 & 0.67 & 0.65 & 0.249 &
 254.0 & 0.3 & 0.62 \\
 {\bf 136}$^2$ & 13.43 & 12.63 & 12.33 & 0.74$\pm$0.04 & 0.75$\pm$0.04 &
 0.71$\pm$0.04 & 0.70$\pm$0.03 & 214$\pm$1 & 0.2 & 0.59 \\
 {\bf 139}--140$^{\dagger2}$ & & 9.17 & & & 0.26$\pm$0.04 & & 1.23$\pm$0.03 &
 86$\pm$1 & 1.4 & 0.83 \\
 {\bf 144}--{\bf 143} & 9.99 & 9.21 & & 1.01 & 0.73 & & 5.371 & 338.3 &
 1.4 & 0.60 \\
 {\bf 157}--{\bf 158} & 10.32 & 9.52 & 9.01 & 0.95 & 0.72 & 0.46
 & 4.321 & 101.0 & 0.5 & 0.60 \\
 166$^\dagger$ & 9.09 & 8.33 & 8.11 & 2.04 & 1.49 & 1.24 & 0.559 &
 151.3 & & 0.35 \\
 {\bf 184}$^3$ & 9.95 & 9.07 & 8.70 & 2.55 & 1.84$_{-0.10}^{+0.50}$ &
 2.08 & 0.409 & 349.7 & 0.4 & 0.27 \\
 {\bf 192}$^1$ & & 12.97 & 12.41 & & 1.35$\pm$0.05 & 0.87$\pm$0.05 &
 0.13$\pm$0.01 & 186.6$\pm$0.5 & 0.2 & 0.38 \\
 {\bf 211} & 12.23 & 10.84 & 10.46 & 1.52 & 1.48 & 1.40 & 1.028 &
 155.6 & 0.7 & 0.35 \\
 {\bf 261}--{\bf 104} & & 8.63 & & & 1.46 & & 6.529 & 312.6 & 1.9 &
 0.36 \\
 {\it TJ\,81}$^{\dagger2}$ & 6.76 & 6.53 & 6.51 & 0.20$\pm$0.04 & 0.14$\pm$0.04
 & 0.14$\pm$0.04 & 0.61$\pm$0.03 & 16$\pm$1 & & 0.91 \\
\hline
\multicolumn{11}{c}{Probable background companions and non-member
 primaries}\\
\hline
 20--21 & & 9.63 & & & 1.88 & & 2.129 & 112.4 & & \\
 100$^1$ & 14.63 & 13.71 & 13.49 & -- & -- & 0.62$\pm$0.05 & 0.13$\pm$0.01 &
 100.7$\pm$0.5 & & \\
 {\bf 124}$^4$ & & 10.92 & & & 5.38$\pm$0.07 & & 2.536 & 103.2 & & \\
 {\bf 137}$^4$ & 10.63 & 9.86 & & 5.86$\pm$0.09 & 5.03$\pm$0.08 & &
 6.734 & 314.5 & & \\
 {\bf LkH$\alpha$86}$^4$ & & 11.03 & & & 7.44$\pm$0.09 & & 2.588 &
 339.2 & & \\
\hline
\end{tabular}

$^1$ deconvolved images -- $^2$ AO locked on a binary system -- $^3$
poor PSF fitting in $H$ -- $^4$ background companion -- $^\dagger$
stars identified as members by L98.
\end{table*}

\begin{table*}[t]
\caption{\label{singles} Stars that are unresolved in our survey;
 boldface entries are for H98's members.}
\begin{tabular}{rrrrrrrrrrrrrrrrr}
\hline
\multicolumn{17}{c}{IfA numbers of single stars}\\
\hline
{\bf 14}&41&{\bf 43}&{\bf 57}&61&67&{\bf 70}&{\bf 78}&{\bf 80}&{\bf 83}&
{\bf 89}&{\bf 93}&{\bf 94}&{\bf 103}&{\bf 106}&{\bf 107}&{\bf 114}\\
{\bf 116}&{\bf 118}&{\bf
  121}&126$^\dagger$&127$^\dagger$&128$^\dagger$&{\bf 131}&{\bf
  134}&{\bf 142}&{\bf 145}&{\bf 146}&147&{\bf 148}&{\bf 152}&{\bf
  154}&{\bf 155}&156$^\dagger$\\
{\bf 159}&{\bf 160}&{\bf
  163}&165&{\bf 167}&169&{\bf 170}&{\bf 171}&{\bf 173}&{\bf 178}&{\bf
  179}&{\bf 181}&{\bf 182}&{\bf 183}&{\bf 185}&186$^\dagger$&{\bf
  187}\\
{\bf 190}&191$^\dagger$&193&{\bf 197}&{\bf 205}&{\bf 206}&{\bf
  210}&220$^\dagger$&{\bf 252}&253&{\bf
  254}&255&\multicolumn{2}{c}{{\bf
  LkH$\alpha$96}}&\multicolumn{2}{c}{{\bf LkH$\alpha$97}}&\\
\multicolumn{2}{c}{{\bf LkH$\alpha$98}}&\multicolumn{2}{c}{{\bf
    LkH$\alpha$100}}&\multicolumn{2}{c}{\it TJ\,89$^\dagger$} &&&&&&&&&&&\\ 
\hline
\end{tabular}

$^\dagger$ stars identified as members by L98.
\end{table*}

Fig.\,\ref{limit} shows the magnitude difference in the $H$ band as a
function of separation for detected binaries. The solid and dotted
lines show the detection limit of our survey, estimated in two
different ways: the solid histogram was established by adding faint
stars around single targets and visually inspecting the images. Since
it corresponds roughly to a 5\,$\sigma$ peak detection, it lies 0.5 to
1\,mag above the dotted line, which represents the 3\,$\sigma$ noise
level as measured on the PSFs of single stars in our images. At large
distances from the primaries ($>1''$), companions can be detected down
to $\sim6.5$\,mag fainter than the primaries. In some cases, even
somewhat fainter stars can be detected. The fainter stars observed in
the mosaics on the first night were observed with smaller
signal-to-noise ratios. The detection limit at large separations is
thus poorer for these stars than in Fig.\,\ref{limit}; close to the
primaries, however, the detection limit remains roughly unchanged,
since the limitation comes primarily from photon noise.

Despite a large dynamic range in our images, we found only 3
secondaries with $\Delta H>2.5$, with in fact $\Delta H>5$. The
location of the widest of these companions (marked "back." in
Fig.\,\ref{limit}) in a $J$--$(J-H)$ diagram indicates that it is a
background star, lying well away from all known cluster members. The
two other very faint companions (marked "back.?") are also likely
background stars, although we lack multicolour photometry to prove it.
For the binary IfA\,139--IfA\,140, we only have $H$ photometry
available; we believe it is a physical binary, however, because of its
close separation. The membership of IfA\,140 could not be determined
by H98 since no $I$ band photometry was obtained for this star; L98
classify it as a member. Similarly, IfA\,82 lacks $V$ measurement in
H98's study, so that its membership is not decided.  We consider,
however, the pair IfA\,85--IfA\,82 as a physical binary, because of
the late spectral type of IfA\,82 (M4, H98). For all other systems,
the location of the companions in a $H$--$(H-K)$ diagram suggests
membership and thus physical association with the primary.

\begin{table*}[t]
\caption{Completeness correction for this survey. ``fraction missed''
  is the ratio of the number of undetected companions to the total
  number of companions. The last column summarizes the overall
  figures. Only H98 members are considered here.}\label{compl}
\begin{tabular}{lccccccc|c}
\hline
sep. range ($''$) & 0.1--0.2 & 0.2--0.3 & 0.3--0.5 & 0.5--0.75 & 0.75--1.0 &
1.0--2.0 & 2.0--8.0 & 0.1--8.0 \\
orbital period (log(d)) & 5.0--5.4 & 5.4--5.7 & 5.7--6.0 & 6.0--6.3 & 6.3--6.5
& 6.5--6.9 & 6.9--7.8 & 5.0--7.8 \\
$\Delta H_{lim}$ (mag) & 2.0 & 2.5 & 3.0 & 4.0 & 5.0 & 6.0 & 6.5 & \\
$q_{min}$ & 0.24 & 0.17 & 0.12 & $<0.1$ & $<0.1$ & $<0.1$ & $<0.1$ & \\
fraction missed & 22\% & 9\% & 3\% & -- & -- & -- & -- & 4\% \\
detected companions & 1 & 2 & 1 & 1 & 0 & 3 & 4 & 12 \\
corrected companions & 1.3 & 2.2 & 1.0 & 1 & 0 & 3 & 4 & 12.5 \\
\hline
& \multicolumn{3}{c}{$\underbrace{\hspace*{3.5truecm}}$} &
\multicolumn{3}{c}{$\underbrace{\hspace*{3.5truecm}}$} & & \\
IC\,348 binary fraction (\%) & \multicolumn{3}{c}{6.8$\pm$3.4} &
\multicolumn{3}{c}{6.1$\pm$3.0} & 6.1$\pm$3.0 & 18.9$\pm$5.3 \\
G-dwarf fraction (\%) & \multicolumn{3}{c}{10.6$\pm$1.2} &
\multicolumn{3}{c}{7.4$\pm$0.8} & 5.2$\pm$0.6 & 23.1$\pm$2.6 \\
\hline
\end{tabular}
\end{table*}

Close to the primaries, it is difficult to detect very faint
companions, because of the wings of the PSF: below $0\farcs2$, only
binaries with flux ratios $\Delta H<2$ can be resolved. To calculate
the actual binary fraction in IC\,348, we need to estimate the number
of fainter, undetected secondaries at these separations. The method we
use, fully described in Bouvier et al. (\cite{bouvier:1997}) for their
survey of the Pleiades, consists in estimating the detection limit in
several separation bins (chosen such that it is roughly constant
within each bin); this flux ratio limit is then converted into a
limiting mass ratio using the mass-luminosity relationship for
2\,Myr-old stars from Baraffe et al. (\cite{baraffe:1998}). Finally,
it is assumed that the mass ratio distribution observed by DM91 in the
solar neighbourhood for G-dwarf binaries applies to IC\,348 binaries.
Then, the limiting mass ratio can be transformed into a fraction of
missed companions. Because of the young age of the cluster: a mass
ratio of $q=0.1$, which is the limit of the DM91 survey, corresponds
to $\Delta H\sim3$\,mag at 2\,Myr. It can be seen, from
Fig.\,\ref{limit}, that we reached this flux ratio over almost the
entire range of separation considered. For the innermost $0\farcs3$,
where the limiting flux ratio of our observations is slightly smaller,
companions close to the $q=0.1$ limit remain undetected. Therefore,
the overall correction is small, with only about 4\% of the companions
missed. We estimate the number of such companions per bin of
separation in Table\,\ref{compl}.  The mass ratio distribution of
M-dwarfs may be flatter than that of G-dwarfs (Fischer \& Marcy
\cite{fischer_marcy:1992}; Reid \& Gizis \cite{reid_gizis:1997a}),
leading to a different estimation of the completeness correction.
However, a flatter distribution implies that we have missed even less
companions (the number of binaries with small $q$ is smaller), so that
our estimate of the number of missed companions can be considered as a
conservative upper value.

In Table\,\ref{compl}, we also estimate the MS binary fraction
("G-dwarf fraction") in different separation ranges by integrating the
binary distribution from DM91. This is the number of binaries as a
function of the orbital period. For IC\,348 binaries, we only have the
angular separation between both components. Therefore, we need to
convert these separations into orbital periods. We use the distance to
the cluster and a statistical correction for the projection of the
semi-major axis on the sky ($\overline{\log a} = \overline{\log \rho}
+ 0.1$, where $a$ is the actual semi-major axis and $\rho$ is the
apparent separation, Reipurth \& Zinnecker
\cite{reipurth_zinnecker:1993}). We also assume that the average total
mass of a system is 1$M_\odot$ (mean value for the observed binaries);
a typical mass of 5$M_\odot$ was assumed for BD+31$^\circ$643. Small
changes in the assumptions about the distance to the cluster, the
projection correction, and the stellar masses do not change the
results significantly, since the MS orbital period distribution is
very broad: a small shift in the integration boundaries does not
significantly modify the binary fraction in this range.

The overall binary fraction (number of companions per observed target)
in IC\,348 is $19\pm5$ \% in the separation range 40--3200\,AU; here,
linear separations have been corrected to account for projection
effects. It is not modified if L98's members are included. From DM91,
we evaluate that the MS G-dwarf binary fraction over the same range is
$23\pm3$~\%.  M-dwarfs have a binary fraction of about 18\% over the
same separation range (Fishcer \& Marcy \cite{fischer_marcy:1992}),
again very similar to IC\,348. We checked that the presence of a few
stars with early spectral type (earlier than K0) in our sample does
not bias our results: the binary fraction in the subsample of stars
with spectral type later than K0 is $17\pm7$ \%, indistinguishable of
that of the whole sample.

\begin{figure}[t]
\label{distrib}
\begin{center}
\includegraphics[height=\columnwidth,angle=270]{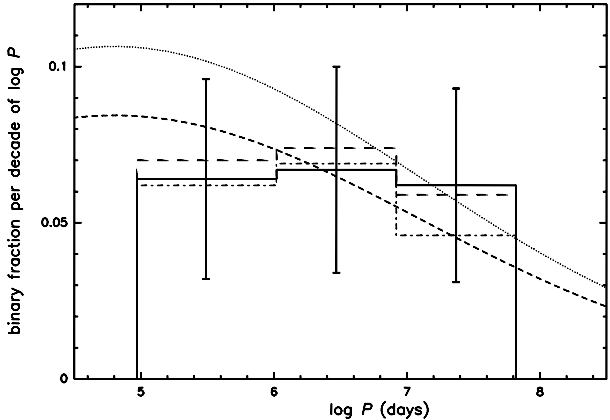}
\end{center}
\caption{Orbital period distribution in IC\,348 (solid histogram),
  compared to the empirical distributions for G- (dotted curve, DM91)
  and M-type (dashed line, Fischer \& Marcy \cite{fischer_marcy:1992})
  MS stars. The long-dashed histogram includes L98's members, while
  the dotted-dashed histogram represent the subsample of stars with
  spectral type later than K0. For clarity, error bars are only drawn
  for one histogram.}
\end{figure}

The orbital period distribution of IC\,348 binaries is shown in
Fig.\,\ref{distrib}. The comparison with the MS for each bin of the
histogram is also given in Table\,\ref{compl}. Both the plot in
Fig.\,\ref{distrib} and the similar values for the binary fraction in
the three separation ranges in Table\,\ref{compl} demonstrate that the
observed distribution is rather flat and, within the errors, not
different from the MS distribution. It is again noticeable that the
results are not strongly modified if we include or exclude higher mass
stars or L98's members. We will now only consider H98 members, for
which masses and ages have been determined.

To evaluate the impact of possible background stars, we estimated the
binary fraction over a smaller separation range, with an upper limit
set at 2$''$ (640\,AU). In this case, the binary fraction for IC\,348
and the MS are respectively $13\pm5$ \% and $19\pm2$ \%, i.e. similar
values, possibly with a small deficiency in IC\,348. This suggests
that the number of false detections is small in the separation range
we selected. On the other hand, in the range 8--16$''$, we find 6
``companions'', that is a binary fraction of $9\pm4$ \%, while the MS
value is $1.8\pm0.2$ \%. Furthermore, the only companion in this
separation range for which we have two-colour photometry appears to be
a background star. This supports our choice for the upper limit:
larger values would imply a non-negligible background star
contamination (unless the period distribution were different from that
in the MS, with a peak at much larger separations). Using the stellar
density in the $H$ off-field image from Lada \& Lada
(\cite{lada_lada:1995}) without any correction for extinction (i.e.,
overestimating the number of background companions), we expect about 4
false detections, similar to our findings. On the other hand, one has
to determine the occurence of projection pairing of two members. The
projected stellar density of the cluster can be crudely estimated from
Lada \& Lada (\cite{lada_lada:1995})'s survey. Once field star
contamination is substracted, they are left with an average density of
$2.5\,10^{-4}$ stars/$''^2$, implying a total of about 3 members
chance projection in the range $0\farcs1$--8$''$. The pair
IfA\,261--IfA\,104 is a good candidate for such a projection effect,
since two other members (IfA\,106 and 107) lie within 15$''$ away from
IfA\,261. We do not try to correct for this effect since we lack a
local estimate of the stellar density around each binary candidate.

\section{Binary properties in IC\,348}
\label{properties}

Various properties of stars, especially of PMS objects, depend on the
stellar environment. The presence of a companion in the vicinity of a
star modifies this environment in a non-negligible way. Potentially,
this can affect the physical properties of the stars in multiple
systems. In Sect.\,\ref{multiple} we first consider the fact that we
have not found any triple or higher order multiple system. Then, the
activity properties of binaries are compared to those of single stars
(Sect.\,\ref{activity}). Finally, our non-detection of very small mass
ratios ($q<0.25$) is discussed in Sect.\,\ref{lowmass}, as well as the
absence of candidate brown dwarfs.

\subsection{High order multiple systems}
\label{multiple}

In our survey, we have found 12 binaries, but no triple or quadruple
systems. Does this mean that there is a deficiency of higher order multiple
systems in IC\,348?

In the solar neighbourhood, G-dwarfs host roughly 10 binaries for 1
triple system, and 4 triple systems for each quadruple system (DM91).
Systems with more than two stars are thus quite rare. For Taurus PMS
stars, Leinert et al. (\cite{leinert:1993}) find a ratio of binaries
to higher order multiples of about 8:1, showing that the number of
triples and quadruples does not seem to evolve significantly from PMS
to MS stages. If we assume that this ratio of 10 to 1 is also relevant
for IC\,348, then we would have expected to find one triple system,
which is not statistically different from our findings, given the
small numbers involved.

There may be a second reason why we failed to detect triple systems.
These systems are usually hierarchical (in both MS and PMS
populations), with a close system surrounded by the orbit of a third
star lying further away. Usually, the ratio of the two semi-major axes
in triple systems is at least $\approx$5 (Tokovinin
\cite{tokovinin:1997}). Given the distance to IC\,348, the peak of
the orbital period distribution corresponds roughly to the smallest
separation we can resolve (see Fig.\,\ref{distrib}), and most of the
triple systems that have been detected in other SFRs have their orbits
on both sides of this peak. Therefore, in IC\,348, triples may just
appear as wide binaries, with the close binary system remaining
undetected.

The absence of any triple system from our sample most likely is not an
indication for a different binary-to-triple ratio between IC\,348 and the MS;
it is probably due to the distance to the cluster and to statistical
uncertainties.

\subsection{Stellar activity in binary systems}
\label{activity}

Several indicators of T\,Tauri star activity have been identified so
far. X-ray and Balmer line (e.g., H$\alpha$) emission are some of
these indicators. We have searched in our sample for a possible impact
of binarity upon this activity (see Table\,\ref{emission}). Although
the samples are rather small, it appears that binaries and single
stars have roughly the same fraction of emitting stars.

It seems, from our survey, that the magnetic activity of PMS stars is
not dependent on the presence of a companion. Binary surveys in X-ray
selected samples had to face the bias induced in their target
selection by the fact that both stars can be emitters. Various
estimates of this bias were obtained by Brandner et al.
(\cite{brandner:1996}) and K\"ohler \& Leinert
(\cite{kohler_leinert:1998}), but it appeared that it is not an
important effect. This is confirmed by the absence of a significant
difference in the binary fraction of $ROSAT$-detected and undetected
stars in a single cluster (Leinert et al. \cite{leinert:1993};
K\"ohler \& Leinert \cite{kohler_leinert:1998}). Similarly, in
IC\,348, the X-ray source sample does not show a higher binary
frequency than the whole sample.

\begin{table}[t]
\caption{\label{emission} Comparison of the overall and binary samples 
 regarding their X-ray and H$\alpha$ emission. A binary is considered as 
 emitting if at least one of its components shows emission.}
\begin{tabular}{llcc}
 & & observed & binaries \\
\hline
 $ROSAT$ & sources & 30 & 6 \\
 & undetected & 36 & 6 \\
\hline
 H$\alpha$ & emission & 31 & 6 \\
 & unknown or abs. & 35 & 6 \\
\hline
\end{tabular}
\end{table}

Because H$\alpha$ emission is likely linked to the accretion
phenomenon on T\,Tauri stars (e.g., Edwards et al.
\cite{edwards:1994}), it also appears that binary members in IC\,348
are surrounded by accretion disks in the same proportion as single
stars, i.e., that binaries do not disrupt disks more rapidly than
singles. Indeed, if two stars are separated by a few tens of AU, inner
disks can remain unaffected around these stars. This is enough to
support accretion onto the stars and to emit Balmer lines. Similarly,
Prato \& Simon (\cite{prato_simon:1997}) showed that the near-infrared
emission of accretion disks is the same in multiple systems and in
single T\,Tauri stars in the Taurus-Auriga SFR, provided the
companions are separated by 40 AU or more.

\subsection{Binary mass ratios and very low-mass companions}
\label{lowmass}

An estimate of the mass ratio for each binary candidate is given in
Table\,\ref{photom}; it has been obtained from the $H$ band relative
photometry, using the 2\,Myr mass-luminosity relationship from Baraffe
et al. (\cite{baraffe:1998}), which can be approximated by
$M_H=-3.25\log M +2.19$ for low-mass stars ($M\leq1M_\odot$), and
assuming that both stars are coeval and equally extincted. Because of
the time dependency of the mass-luminosity relationship, these
estimates are somewhat uncertain. We used the median age determined
by H98 as a typical value for the whole cluster. Also, the extinction
along the line of sight of the primary and the secondary are unknown
and might be different (this effect should be rather small at
1.65\,$\mu$m, however), and infrared excesses can represent a
significant part of the flux at this wavelength. In two cases
(IfA\,144 and 157), the mass ratio estimated from $\Delta J$ is
significantly smaller than from $\Delta H$, indicating that at least
one component shows a significant excess (IfA\,143 and 158 secondaries
actually show larger $J-H$ excesses than their primaries by about
0.2\,mag); in the other cases, both values are similar. Also, the $H$
band relative photometry for IfA\,184 is somewhat uncertain. However,
we assume that these mass ratios are not systematically biased towards
low or high values.

We did not find any companion fainter than $\Delta H\approx2$\,mag,
although it would have been easily detected, provided that the
separation of the system is larger than $0\farcs2$ (see
Fig.\,\ref{limit}). This flux ratio corresponds to a mass ratio of
$q=0.25$ at 2\,Myr. For an average primary mass of 0.5$M_\odot$, this
could point to the absence of very low-mass stars and brown dwarfs as
secondaries. Alternatively, a statistical fluctuation cannot be
excluded given the small number of detected binaries.

\begin{figure}[t]
\begin{center}
\includegraphics[height=\columnwidth,angle=270]{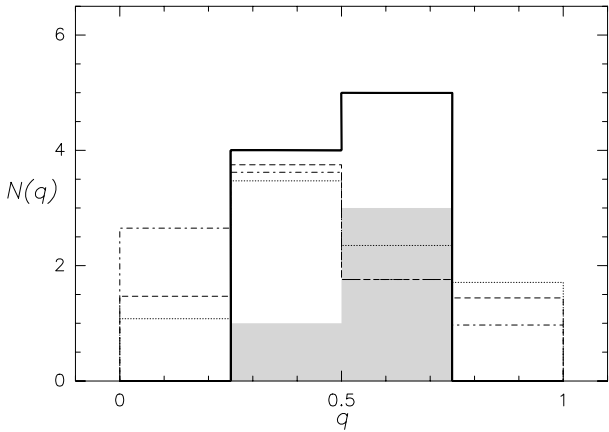}
\end{center}
\caption{\label{ratio} Mass ratio distribution for binaries in IC\,348
  with $M_A\leq1M_\odot$ ({\it thick histogram\/}) compared to
  distributions simulated by taking at random the masses of the
  companions from some IMFs. {\it dotted\/}: Kroupa et al.
  (\cite{kroupa:1993}) with $\alpha_1=1.3$; {\it dashed\/}: Reid \&
  Gizis (\cite{reid_gizis:1997a}) in their 5.2\,pc sample, without
  brown dwarfs; {\it dotted-dashed\/}: Reid \& Gizis
  (\cite{reid_gizis:1997a}) with brown dwarfs down to 0.05\,$M_\odot$
  and with a flat mass distribution in the substellar domain. The
  latter model predicts a high number of binaries in the first bin,
  contrasting with the observations; it is very similar to the
  observed mass ratio distribution in the MS (DM91). The shaded
  histogram represents the subsample of low-mass stars
  ($M<0.3M_\odot$) in IC\,348.}
\end{figure}

In order to test the significance of this result, we performed
Monte-Carlo simulations to compare the observed mass ratio
distribution in IC\,348 with that obtained assuming that each
companion has a mass (lower than its primary) taken at random from a
given initial mass function (IMF). Each simulated histogram is the
average of 1000 simulations, so that statistical uncertainties can be
neglected; the exclusion of binaries with $q<0.1$ does not modify the
histograms by more than a few hundredths in each bin. First, we used
the Kroupa et al. (\cite{kroupa:1993}) IMF, with $\alpha_1=1.3$ and
stellar masses in the range 0.08--1$M_\odot$. The number of predicted
binaries in the range $q=0$--0.25 is about one. We also used the Reid
\& Gizis (\cite{reid_gizis:1997a}) IMF within the 5.2\,pc solar
neighbourhood, with and without brown dwarfs (i.e., with a minimum
mass of 0.05 and 0.075$M_\odot$ respectively). In the first case, the
mass function was chosed flat in the brown dwarf domain
($\psi(M)\propto M^0$), following Reid \& Gizis
(\cite{reid_gizis:1997b}); we checked that the slope of the mass
funtion does not modify significantly the results. Without brown dwarf
companions, we again predict about one companion in the first bin of
Fig.\,\ref{ratio}, while this number is increased to almost three if
we include substellar objects. In all three cases, the number of
detected companions in the range $q=0.5$--0.75 is about twice as large
as that predicted by our simulations. In Fig.\,\ref{ratio}, only
binaries with $M_A\leq1M_\odot$ are plotted, because Reid \& Gizis's
IMF is only defined below 1$M_\odot$. It should be noted that the
random pairing assumption together with the latter IMF leads to a
mass-ratio distribution significantly different than that observed by
Reid \& Gizis (\cite{reid_gizis:1997b}).

Despite of the small size of our small binary sample, the observations
indicate that binaries in IC\,348 are not preferentially equal mass
systems, since no binary appears in the bin $q=0.75$--1.
According to DM91, the mass ratio distribution for solar-type field
stars peaks near $q=0.2$--0.3. On the other hand, in their studies of
low-mass stars in the solar neighbourhood and the Hyades cluster, Reid
\& Gizis (\cite{reid_gizis:1997a,reid_gizis:1997b}) concluded that
M-dwarf binaries have a mass ratio distribution peaking at
$q\approx1$, which contrasts with our findings fot IC\,348. It is
unlikely that our observations have missed some equal mass systems;
unless infrared excesses introduce a systematic bias against equal
flux binaries, this deficit is real.

The absence of small mass ratios ($q<0.25$) in our observations is
only marginally significant, given the sample size. It is however more
consistent with the models without brown dwarf companions.  Reid \&
Gizis (\cite{reid_gizis:1997b}) concluded from their Hyades study that
the mass function is flattening, maybe even decreasing, in the
substellar domain. However even a flat mass function probably predicts
too much companions with $q<0.25$ in IC\,348. It rather seems that
{\it no\/} brown dwarfs are companions of low-mass stars. This absence
of brown dwarfs is also found in Table\,\ref{photom}, where all
companions have stellar masses, with the possible exception of the
companion of IfA\,192 ($M_B\approx 0.075M_\odot$). We caution again
that these estimates are somewhat uncertain, due to age effects,
infrared excesses and, in some cases, unknown extinctions; spectra of
the candidates should be obtained to determine their
stellar/substellar status (IfA\,49, 82, and 184B have masses below
0.15\,$M_\odot$ from our results).

If this absence of brown dwarfs is not due to statistical fluctuations
or to systematic errors in the estimate of the mass ratio, this
suggests that brown dwarfs cannot form in IC\,348 at separations
larger than $\approx50$\,AU from stars. The results of Reid \& Gizis
(\cite{reid_gizis:1997b}) indicate that this limit is smaller than
5\,AU in the Hyades. This apparent lack of very low-mass companions
may result from dynamical biasing during the early evolution of small
subclusters: $N$-body simulations of Sterzik \& Durisen
(\cite{sterzik_durisen:1998}) show that, in most cases ($\sim90$\,\%),
the dynamical evolution of small-$N$ systems result in the association
of the two more massive stars in a binary system and to the ejection
of the lower mass components.  However, in the solar neighbourhood,
very low-mass secondaries ($M<0.1M_\odot$) can be found, at
separations varying from 4 to 1800\,AU (Reid \& Gizis)
\cite{reid_gizis:1997a}).

\section{Environmental conditions and binary formation}
\label{discus}

In order to investigate evolutionary effects, we first compare the binary
fraction we have determined in Sect.\,\ref{fraction} with that of other SFRs
(Sect.\,\ref{compar}). The possible temporal evolution of the binary fraction
is discussed in Sect.\,\ref{environ}, before we argue that environmental
conditions and binary frequency may be tightly linked.

\subsection{Comparison of binarity in IC\,348 with other SFRs}
\label{compar}

Given that the stars in IC\,348 have a median age of 2\,Myr, we can
compare them directly with other SFRs like the Taurus-Auriga complex
and the Orion Trapezium cluster, without introducing an evolutionary
bias in terms of age. The binary excess observed in Taurus,
Cham{\ae}leon and Ophiuchus is of the order of a factor of 1.6
(Duch\^ene \cite{duchene:1998bin}). A similar excess, if existing in
IC\,348, would yield a binary fraction of $37\pm5$ \% in our surveyed
separation range. This is different from our observed binary fraction
at a 2.5\,$\sigma$ level, and can be excluded with a high confidence
level ($>$98\,
likely harbor different binary fractions. From this we conclude that
not all SFRs have a unique binary fraction, several of them exhibiting
excesses (Taurus, Ophiuchus, Cham{\ae}leon) and others showing binary
fractions similar to that of the MS (the Trapezium cluster, IC\,348,
as well as other clusters in Orion: NGC 2024, 2068 and 2071, Ghez et
al. \cite{ghez:1997}). Consequently, we are led to the conjecture that
the age, which is the same on average for the above-mentioned SFRs, is
not the only parameter governing the binary fraction in a young
cluster or a T-association.

The stars we observed in IC\,348 represent a rather large age spread,
from a few $10^5$ to about $10^7$ years. This allows a comparison of
the binary fraction with stellar age, which is shown in
Table\,\ref{age}. We selected all stars in our samples with known age,
which excludes the stars only detected in the H$\alpha$ survey, as
well as two other members which lack $I$ photometry in H98's study
(IfA\,83 and 89). It appears that all three subsamples have similar
binary fractions, although we are limited by the small sample sizes; a
rank order test does not indicate any difference between the single
and binary stars age distributions. This indicates that the binary
fraction does not evolve significantly with time, at least on a
timescale of a few Myrs. We also verified that the binaries in the
three subsamples defined in Table\,\ref{age} are roughly equally
represented at all separations.  This means that we do not see any
indication for an evolution of the binary separations over the
timescale of the age range covered by these stars.

\begin{table}[t]
\caption{\label{age} Evolution of the binary fraction with primary age, as
  estimated by H98. There are a few stars in our sample with no age
  estimate (IfA\,83 and 89, and the H$\alpha$ stars).}
\begin{tabular}{cccc}
 & $t<10^6$ & $10^6<t<10^{6.5}$ & $t>10^{6.5}$ \\
\hline
 observed & 20 & 22 & 16 \\
 binaries & 5 & 3 & 4 \\
\hline
\end{tabular}
\end{table}

As an alternative to an evolutionary process, it has been proposed
that the binary excess observed in various SFRs was the result of an
observational bias: since the mass-luminosity relationship is
shallower for younger stars, it is easier to detect companions around
PMS stars than in the MS population (Zinnecker, priv. com.). The fact
that we know several SFRs, now including IC\,348, with no binary
excess indicates that this bias is not responsible for the observed
overabundance of PMS binaries in some regions.

The binary fraction that seems to differ between SFRs relates to
binaries which cover only a limited separation range. If the
separation distribution was different from one SFR to another, the
overall binary fractions could still be the same for all of them: an
excess observed in a given separation range could be balanced by a
deficiency of binaries with shorter or longer periods. There is
currently no such evidence, except perhaps for the study of the
$ROSAT$ population in Upper Scorpius by Brandner \& K\"ohler
(\cite{brandner_kohler:1998}). Several arguments indicate that the
orbital period distribution does not vary significantly between PMS
and MS binaries: the number of spectroscopic binaries in Taurus is at
least as large as that of the MS (Mathieu \cite{mathieu:1994}), lunar
occultation surveys in this SFR have shown that the binary excess was
present down to $\sim$1\,AU (Richichi et al. \cite{richichi:1994};
Simon et al. \cite{simon:1995}, and Pleiades binaries have a similar
period distribution as dwarfs (Mermilliod et al.
\cite{mermilliod:1992}; Bouvier et al. \cite{bouvier:1997}).

\subsection{Binary fraction and environmental conditions}
\label{environ}

Considering IC\,348, the Trapezium (Prosser et al.
\cite{prosser:1994}; Petre et al. \cite{petr:1998}) and Pleiades
(Bouvier et al. \cite{bouvier:1997}) clusters, and the solar
neighbourhood stars (DM91), we have four samples with no binary excess
at different evolutionary stages (PMS, ZAMS and MS). We thus conclude
that the binary fraction does not evolve with time between these
stages. Any evolution of the binary frequency would have to occur
within the first Myr after the formation process. Furthermore, the
differing binary fractions between the various SFRs of the same age
have to be explained, and a global time effect cannot be responsible
for this.  One common property of all the clusters without binary
excess is that they are all rather dense: IC\,348 has about 500
stars\,pc$^{-3}$, and the Trapezium is about 10 times denser. The
older Pleiades cluster, which is still dense nowadays, was probably
even denser when younger, perhaps similar to the Trapezium. On the
other hand, the SFRs with high binary fractions (Taurus, Ophiuchus,
Cham{\ae}leon) are rather loose, with no more than a few
stars\,pc$^{-3}$ in the Taurus aggregates. This seems to indicate that
a link exists between the binary fraction and the cluster density.

Several physical processes could be the reason behind such a link. The
impact of the average cluster density on the binary fraction could be
direct; for instance, in dense clusters, the number of gravitational
encounters is high and the binaries could be massively disrupted in
such clusters over short timescales. From $N$-body simulations, Kroupa
(\cite{kroupa:1995a,kroupa:1995b}) has shown that in clusters as dense
as the Trapezium cluster the binary fraction could decrease from 100\%
to about 50\% in less than 1\,Myr. Then, a model where all SFRs form
with a high binary fraction (i.e., close to 100\%), and where
gravitational interactions between multiple systems are responsible
for the decrease of the number of binaries, would be in qualitative
agreement with the observational results: in all dense clusters, the
binary fraction would have already decreased down to the MS level even
for the youngest clusters in which binary fractions have been measured
so far, while it would have remained high in loose PMS associations.

Alternatively, it is possible that the density is not the main
parameter governing the binary fraction, but that another physical
parameter, during or even before the star formation process, drives
the subsequent evolution of the cluster, including its stellar density
and binary fraction simultaneously. Durisen \& Sterzik
(\cite{durisen_sterzik:1994}) have shown that the current models of
fragmentation and disk instability may imply higher binary fractions
when the initial cloud temperature is lower. In general, cold giant
molecular clouds may not be very efficient in forming stars, if the
output of their fragmentation is small aggregates, with low densities
(like Taurus). Regions creating high-mass stars, on the other hand,
have rather high cloud temperatures; they usually from dense clusters,
such as the Trapezium cluster. The link we find between binary
frequency and cluster density could then be an intrinsic output of the
fragmentation process. Other characteristics of the cloud before star
formation occurs could as well be responsible for the observed linked
between density and binary fraction. For instance, the nature of the
pre-collapse equilibrium in the parent cloud may influence the mass,
size and angular momentum of the fragmented cores, leading to
differing binary fraction and cluster density.

Environmental conditions at the time of star formation thus could have
an impact on the resulting binary population (i.e., the total number
of multiple systems). At the present time, it is not possible to
distinguish between a very rapid temporal evolution of the binary
fraction, during the first Myr, or an intrinsic dependence of the
binary fraction on these conditions. Observations of even younger
populations in embedded clusters, as well as determination of accurate
orbital period and mass ratio distributions are needed to go further
into the history of binary formation and evolution.

\section{Conclusions}
\label{concl}

From a high-angular resolution study of IC 348 low-mass members we
find that the binary frequency in this very young cluster ($\sim
2$\,Myr) is similar to that of the Pleiades ($\sim 100$\,Myr) and of
low-mass field dwarfs ($\sim 1$\,Gyr). We therefore conclude that the
binary frequency does not significantly evolve over time on a
timescale of several 100 Myr.  Instead, it appears that the binary
frequency among low-mass stars is already established at very young
ages, i.e., within $\sim$1\,Myr after the formation process.

In particular, a long-term evolutionary effect cannot be responsible for
the differing binary fractions found in different SFRs: on the one hand,
Taurus and Ophiuchus exhibit binary excesses, on the other, the Trapezium
and IC\,348 clusters don't. Yet, all these regions have similar ages of
$\sim$1--2\,Myr. Furthermore, in the IC\,348 sample, we do not find
evidence for an evolution of the binary fraction or the orbital period
distribution within the age spread of the cluster of about 10\,Myr. A time
evolution of the binary frequency, if any, is thus constrained to occur
within the first 1\,Myr of stellar evolution. After this time, intrinsic
differences exist between SFRs regarding their binary content.

In spite of the large dynamic range of our images, no brown dwarf
companion is found in IC\,348 binaries (to the possible exception of
IfA\,192B). The mass ratio distribution we find is consistent with the
absence of brown dwarf companions to low-mass members of this cluster;
a similar conclusion was drawn for the Hyades cluster. Also, the
mass-ratio distribution is not peaked towards $q=1$, in possible
contradiction with what has been proposed for the solar neighbourhood
M-dwarfs population. Further studies with larger telescopes will allow
a better determination of this distribution.

Comparing the results obtained on IC 348 to similar studies in other
clusters, it appears that the binary fraction may be inversely correlated
with the average cluster density, with dense clusters showing low binary
fractions (similar to field dwarfs), as opposed to the loose T-associations
like the Taurus-Auriga and Cham{\ae}leon complexes where the binary
fraction is larger. On the basis of this qualitative trend, at least two
scenarios may explain the observed differences in binary fractions: either
the formation mechanism always leads to an initially high binary fraction
(of the order of 100\%) and frequent gravitational encounters in dense
clusters disrupt binaries on a timescale of 1\,Myr or less, or specific
initial conditions in the parent molecular cloud, such as the gas
temperature, metallicity, angular momentum , etc..., lead to different
output of the star formation process and govern simultaneously the binary
frequency and the cluster density. High resolution studies of embedded
clusters even younger than those investigated so far are still needed to
settle these issues.

\begin{acknowledgements}
  Numerous comments by J. Eisl\"offel have significantly improved this
  paper. We also thank G. Herbig for reading an early version of this
  work and for providing us with his members list prior to publication
  and mass estimates, as well as I. Baraffe for making her
  evolutionary models available. Observing support from CFHT is also
  gratefully acknowledged, especially J.-L. Beuzit, M.-C. Hainaut and
  D. Woodworth.
\end{acknowledgements}

\end{document}